\documentstyle[prl,twocolumn,aps]{revtex}
\input{epsf}
\begin{document}
\draft
\twocolumn[\hsize\textwidth\columnwidth\hsize\csname @twocolumnfalse\endcsname
\title{Ground state properties of the 2D disordered Hubbard model}

\author{Ga\"etan Caldara, Bhargavi Srinivasan and Dima L. Shepelyansky}

\address {Laboratoire de Physique Quantique, UMR 5626 du CNRS, 
Universit\'e Paul Sabatier, F-31062 Toulouse Cedex 4, France}

\date{April 6, 2000}

\maketitle

\begin{abstract}
We study the ground state of the 
two-dimensional (2D) disordered Hubbard model by means 
of the projector quantum Monte Carlo (PQMC) method.
This approach allows us to investigate the 
ground state properties of this model for lattice sizes up to
$10 \times 10$, at quarter filling, for a broad
range of interaction and disorder strengths.
Our results show that the ground state of this
system of spin-1/2 fermions remains localised in the presence 
of the short-ranged Hubbard interaction.
\end{abstract}
\vskip1pc]

\narrowtext
\section{Introduction}
The electronic transport properties of disordered
systems have been the subject of much investigation
in physics. From the pioneering work of Anderson (1958)\cite{and58}
it is known that in three dimensions (3D)
the eigenstates of a non-interacting 
electron gas in a
random potential become localised at the Fermi energy above a critical value
of the disorder strength $W_c$. In this regime, the eigenstates
decay exponentially in space and hence cannot carry a 
current; thus the system is an insulator. For disorder strength, $W$, lower
than $W_c$, the eigenstates are extended and 
diffusive transport takes place in the system in accordance with Ohm's law.
However, for two-dimensional (2D) systems, it was shown by the 
scaling theory of Abrahams {\it et al}\cite{and79} that 
all states are localised for any  disorder strength.
Thus, it appears that there is no metal-insulator transition
(MIT) for non-interacting electrons in 2D. The properties of non-interacting
electrons in random potentials have since been studied systematically
and the main physical effects have been understood\cite{lee}.
In this context, the experimental observation by Kravchenko {\it et al}\cite{krav94}
of a transition
from insulating to metallic behaviour, as seen in the resistivity
as a function of temperature of the 2D electron gas, 
came as a great surprise to the community. The existence of this
transition from insulating to metallic behaviour as a function
of density has been confirmed by other groups\cite{popovic,canada,yael,alex}.
The experiments were carried out on very high mobility,  
low electron density ($n_s$) samples
which correspond to a regime where the electron-electron
interactions ($E_{ee}$) are much stronger than the Fermi energy
($E_F$), such that the dimensionless parameter $r_s (\approx E_{ee}/E_F)$
lies in the range 5-50.  This indicates the importance of
electron-electron interactions in these systems.
Further, it was shown experimentally that 
the application of an in-plane magnetic field ($B_{p}$) drives the
system insulating\cite{simonian}. Since such a field can only couple to the
spin, this experiment indicates the important role played by the spin degrees of
freedom. While the early experiments have stimulated a spate of
new experimental results, there has been no satisfactory theoretical
explanation of  the phenomenon of the 2D MIT, to date.

The effects of interactions in disordered systems have been studied
from the metallic side in great detail, 
where the interactions are relatively weak\cite{alar}.
The study of interactions in the localised phase were mainly carried out within
the mean-field approximation, which led to a number of important
results, for example, the Efros-Shklovskii gap in the density of states
near the Fermi level\cite{efros}, but could not take into
account quantum interference effects, important in this many-body
system. The investigation of a simple model of two interacting
particles (TIP) in the localised phase showed that short-range
attractive/repulsive interactions can lead to destruction of localisation
and propagation of pairs of particles on a length scale much larger than
the one-particle localisation length\cite{ds94}. Thus, the effects
of interaction on the localised phase are non-trivial and deserve
a detailed study. This, however, is not an easy task. Indeed, even the 
analytical expressions for the matrix elements of the interaction 
in the localised phase are not known\cite{mirlin}, hence,
numerical studies of the problem become important.

Recent numerical approaches to the question have included the
studies of persistent currents by exact diagonalisation
of small 2D clusters\cite{benen} and Hartree-Fock based calculations
without\cite{poilblanc} and with residual interaction\cite{vojta,hfgap}.
These approaches led to some interesting indications
but did not allow the study of sufficiently large systems 
and/or sufficiently many particles. Other approaches based on 
level spacing statistics of many-body states made possible the study
of larger systems and showed the existence of ergodic 
(delocalised) states for low energy excitations, but not at
the ground state\cite{song}. All these studies were carried out for
spinless fermions. Recently, the properties of fermions with
spin on a disordered 2D lattice were investigated using a finite
temperature quantum Monte Carlo (QMC) method\cite{trivedi}. The
temperature dependence of the 
resistivity obtained numerically indicated a transition from
insulating to metallic behaviour for sufficiently strong 
interactions and weak disorder strength. However, these calculations
were carried out at finite temperatures and technical problems 
("fermion sign problem") did not permit the analysis of the
ground state.  This is not completely relevant to the experiments
which were carried out at temperatures much below the Fermi energy\cite{krav3}
and thus require a better understanding of the properties of the
ground state, as in the general scenario of quantum phase
transitions.

To investigate the properties of the ground state of the disordered
interacting fermionic system, we choose the Hubbard model with
site diagonal disorder.
This could be considered as an important first step
on  the way to investigations of more complicated models
with Coulomb interactions, which might be more appropriate
for experiments at low densities.
We study the ground state of our model
on the square lattice by the projector QMC (PQMC) method. 
We use different characteristics  to investigate the extent of
the ground state wavefunction for a broad range of model parameters:
disorder strength ($W$), interaction strength ($U$) and filling factor ($\nu$).
The studies were carried out in the $S_z = 0$ sector, with equal numbers
of particles with up and down spins.
This is thus the first numerical study of the ground state of a disordered,
interacting system of  fermions with spin.

This paper is organised as follows. In the next section, we 
describe  the model and the method used. In the third section
we present our results for the averaged Green function, the charge densities
and the inverse participation ratios, all of which we use to characterise
the ground state. We present a summary of our conclusions in the
last section.

\section{Model and method}

The two-dimensional disordered Hubbard model on a square lattice is
given by
\begin{equation}
\label{hamil}
\begin{array}{l}
H = H_A + H_I  \\=
\Bigl(
 -t\sum\limits_{\langle ij \rangle, \sigma} 
 \hat{a}^{\dagger}_{i,\sigma} \hat{a}_{j,\sigma}
+ \sum\limits_{i,\sigma} 
\epsilon_i \hat{a}^{\dagger}_{i,\sigma} \hat{a}_{i,\sigma} 
\Bigr)
 +  U \sum\limits_{i} \hat{n}_{i\uparrow}\hat{n}_{i\downarrow} 
\end{array}
\end{equation}
where the $\hat{a}^{\dagger}_{i,\sigma}$ ($\hat{a}_{i,\sigma}$) are the 
creation (annihilation)
operators for a fermion of spin $\sigma$ at site $i$ with periodic 
boundary conditions, $\hat{n}_{i\sigma}$
is the number operator for spin $\sigma$ at site $i$, $t$ is 
the hopping parameter, the Hubbard parameter, $U$, measures
the strength of the screened interaction and $\epsilon_i$, the 
energy of site $i$ is a random number drawn from a uniform distribution
$[-W/2, W/2]$, which parametrizes the disorder. The first two terms
represent the Anderson Hamiltonian and the last term represents
the interaction $H_I$.
In the limit $W = 0$, this Hamiltonian reduces to the usual
Hubbard model. The filling factor $\nu = N_p/(2 \times N^2)$, where
$N_p$ is the number of fermions (particles) and $N$ the linear dimension
of the system; thus the total number of sites is $N^2$.

We obtain the ground state properties of this model by the 
PQMC method. The PQMC method was initially developed for the
Hubbard model and has been used to obtain reliable results 
for large lattices\cite{pqmc}. The method can be generalised 
in a straightforward
manner to include random site energies.  We now present 
some details of the calculation for completeness and refer
the reader to the literature for more detailed accounts.

The PQMC method consists in obtaining the true ground state 
$|\psi_{0} \rangle $ of
the Hamiltonian(1) by projection from
a trial wavefunction $|\phi\rangle$
 that is not orthogonal to the true ground
state of the system,
\begin{equation}
|\psi_{0} \rangle
= \lim\limits_{\Theta \rightarrow \infty} { {e^{-
\Theta{\hat H}}
|\phi\rangle} \over
 {\sqrt{\langle \phi| e^{- 2 \Theta{\hat H}} |\phi \rangle}}}.
\label{projec}
\end{equation}
The  trial wavefunction is usually formed from the eigenstates
of the non-interacting Hamiltonian (orbitals filled up to the
Fermi level). In this case, we choose
the eigenstates of the Hamiltonian $H_A$, thus including the 
random potential. To carry out Monte Carlo (MC) simulations
of this quantum Hamiltonian, it is first necessary to map
it onto an effective classical Hamiltonian. Thus, the projection
operator $\exp(-\Theta \hat{H})$
is first Trotter decomposed as
$\Bigl( \exp(- \Delta \tau {\hat H}_{A})
 \exp(- \Delta \tau {\hat H}_{I})  {\Bigr)}^{L}$ with 
$\Theta = \Delta\tau \times L$. This introduces a systematic error 
of order $(\Delta \tau)^2$ due to non-commutation of
$\hat{H}_A$ and $\hat{H}_I$. The interaction is then decoupled by
a discrete Hubbard-Stratonovich (H-S) transformation, by the 
introduction of $N^2 \times L$ Ising-like fields. Since the 
complete summation over these degrees of freedom is too time-consuming
to be practical,
the method reduces to a MC sampling of physical properties, which 
is the second source of error, the statistical error. It is
important to note that during the MC process, each configuration
of Ising spins is assigned a weight, which is interpreted
as a probability. This quantity is positive definite only 
at half-filling for the uniform Hubbard model. The problems that
arise from the non-positive-definite nature of this quantity
are referred to as the 'fermion sign problem' in the literature
and are known to be particularly severe slightly away from half-filling
in finite temperature methods and restrict the lowest 
temperature that can be attained in the simulation 
(in the clean limit).

We have studied system sizes of up to $10 \times 10$ at 
quarter and one-eighth fillings (50 and 25 particles). We carried out extensive checks
on the MC parameters to assure ourselves of convergence, as described
in Ref.\cite{bs1}, but in the presence of disorder. Thus,
we chose $\Theta = 3.0$, with $L = 30$, to have  $\Delta \tau = 0.1$.
This, with the symmetric Trotter decomposition reduces
the systematic error to $ (\Delta \tau )^3  \approx 0.001 $. We 
checked for statistical convergence of our data in several ways.
By varying the number of sweeps after equilibration, we determined
that 1000 MC sweeps are sufficient for equilibration and 
2000 further sweeps for property estimates. We carried out 
measurements until the standard deviations on our values
were of the order of the systematic error. We also tested
our results against results obtained from exact diagonalisations for small system
sizes and the results for the charge density,
$n_i=n_{i\uparrow}+n_{i\downarrow}=\sum_{\sigma}\langle\hat{n}_{i\sigma}\rangle$,
presented in Fig. 1
show good agreement  with the exact results.
Convergence is of course the best for the ground state energy
as compared to other physical quantities,
and we have a relative accuracy of $10^{-3}$ when compared to exact 
calculations. 
As for the effect
of disorder on the sign problem, it was possible to study
the $10 \times 10 $ lattice for $U/t = 6 $ for disorder strengths $W/t$
of up to 7-10. Our measure of the severity of the sign problem is
to consider the quantity f =  1- (number of negative determinants)/
(total number of
determinants). In all the cases considered, we have f = 0.999,
which indicates that the sign problem is under control. 

\begin{figure}
\epsfxsize=3.1in
\epsfysize=2.7in
\epsffile{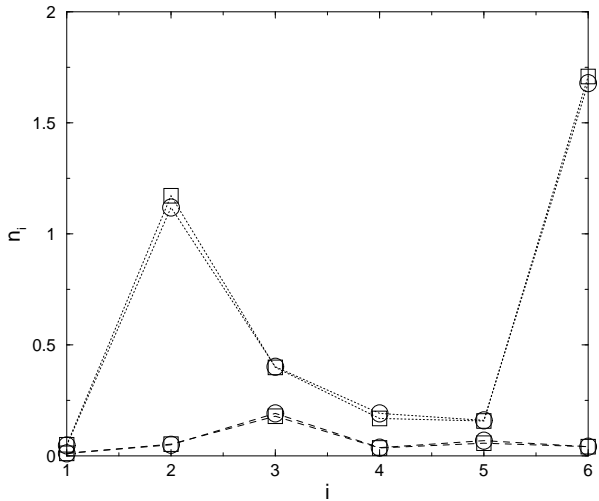}
\vglue 0.2cm
\caption{ Comparison of exact diagonalisation (circles) and PQMC (squares) results for
the charge density ($n_i$) per site ($i$) for a  2-chain Hubbard model,
dotted lines correspond to the upper chain and
dashed lines to the lower chain, 
with system size $6 \times 2$, $U/t = 2, W/t = 10$ and filling at 4 particles.
}
\label{fig1}
\end{figure}


From the simulations, it is possible to obtain ground state 
expectation values of the single particle Green function,
$G_{ij} =\sum_{\sigma} \langle 
\hat{a}^{\dagger}_{i,\sigma} \hat{a}_{j,\sigma} \rangle$,
where the average is a MC average. Further, we can obtain 
ground state expectation values of other one- and two-body
operators, such as the charge density and the charge-charge
correlation functions. Each disorder realisation constitutes
a full PQMC calculation. The properties are
 averaged over 16 disorder realisations. Thus, we have 
obtained the evolution of the Green function with distance,
the charge densities and the inverse participation ratios. Our
results will be described in the following section.

\section{Results and discussion}

To characterise the properties of the ground state, we study the
correlation function defined as
\begin{equation}
C({\bf{r}}) =   \frac{1}{N^2} \langle
\sum\limits_{i,j} | 
\langle \hat{a}^\dagger_{i,\uparrow} \hat{a}_{j,\uparrow} \rangle  +
 \langle \hat{a}^\dagger_{i,\downarrow} \hat{a}_{j,\downarrow} \rangle 
|^2
\delta_{i-j,\bf{r}}\rangle ,
\end{equation}
where ${\bf{r}} = {\bf{i}} - {\bf{j}}$ is the vector in the plane
between the sites labelled $i$ and $j$, 
and the averages are
carried out over 
the ground state eigenfunction and 
the different
disorder realisations for all possible initial positions of 
${\bf{r}}$, i.e. all corresponding $i$ and $j$. 
$C({\bf{r}})$ is simply related
to the Green function $G_{ij}$ already defined in Section II.
With this definition, 
$C(0) = N^{-2} \langle\sum_{i} (n_{i\uparrow}+
n_{i\downarrow})^2 \rangle
\sim 4\nu^2$ in the limit of weak disorder and 
$C(0) \sim 4\nu$ in the strongly localised limit. 
The dependence of $C(r)$ on distance $r$ is related to
the localisation of the eigenstate, 
i.e. we expect exponential decay of this
quantity for localised states and slow decay at long distances 
for extended states. We also study the direction averaged 
correlation function $\bar{C}(r)$ which now depends only on the
distance $r = |{\bf{r}}|$. 

\begin{figure}
\epsfxsize=3.1in
\epsfysize=3.0in
\epsffile{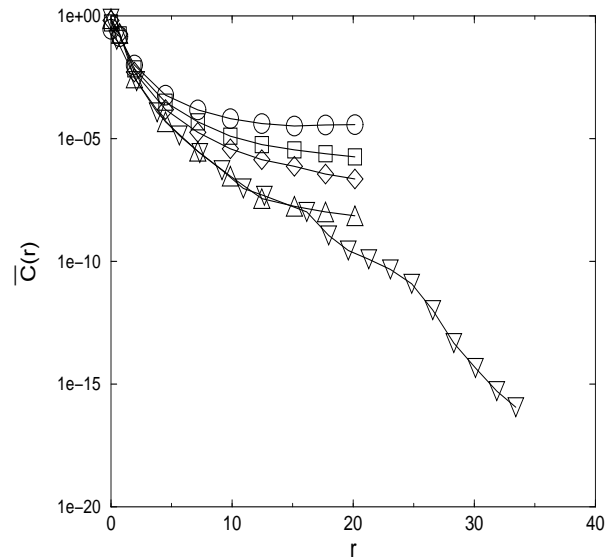}
\vglue 0.2cm
\caption{
 The decay of the direction averaged correlation function
 $\bar{C}(r)$ vs. $r$ for a $30 \times 30$ system for $U/t = 0$, with
 $W/t$ = 2 (circles), 7 (squares), 10 (diamonds), 15 (up-triangles)
 and a $50 \times 50$  system at  $W/t$ = 15 (down-triangles),
 averaged over 16 disorder realisations, at quarter filling.
}
\label{fig2}
\end{figure}

In Fig. 2 we show the decay of $\bar{C}(r)$ with $r$, for the
2D Anderson model, system (1) at $U/t = 0$, for various disorder strengths.
The change from flat behaviour with $r$ at weak disorder,
when the eigenfunctions are delocalised in the finite sized 
system ($W/t = 2$), to asymptotic exponential decay for stronger disorder,
$W/t \geq 10$,
(when the localisation length is smaller than the system size),
is evident. We note that the initial non-exponential decay in
the localised case is due to the fact that the ground state eigenfunction is 
a superposition of one-particle eigenstates of different 
energies. In fact, the one-particle localisation length $l_1$ of
a state depends 
on its energy and therefore the many-body state,
which is the Slater determinant of the 1-particle states up to
the Fermi level, initially decays more rapidly. 
This is due to the low-lying
states of smaller localisation lengths and it is only in the 
asymptotic limit that the decay of the many-body 
ground-state is 
determined by the maximum $l_1$ at $E_F$.
This physical  structure of many-body states complicates the 
observation of asymptotic exponential decay corresponding to the
one-particle localisation length $l_1(E_F)$ at the Fermi level.
Despite these complications, the asymptotic slope 
is seen to depend strongly on $W$ (Fig. 2), which is
consistent with the exponential growth of $l_1$
with decreasing $W$ in 2D\cite{lee}.
In view of this, the investigation of the
correlation function  $\bar{C}(r)$ in the presence of interactions 
should tell us the impact of interactions on the 
localisation properties of eigenstates.

\begin{figure}
\epsfxsize=3.1in
\epsfysize=2.8in
\epsffile{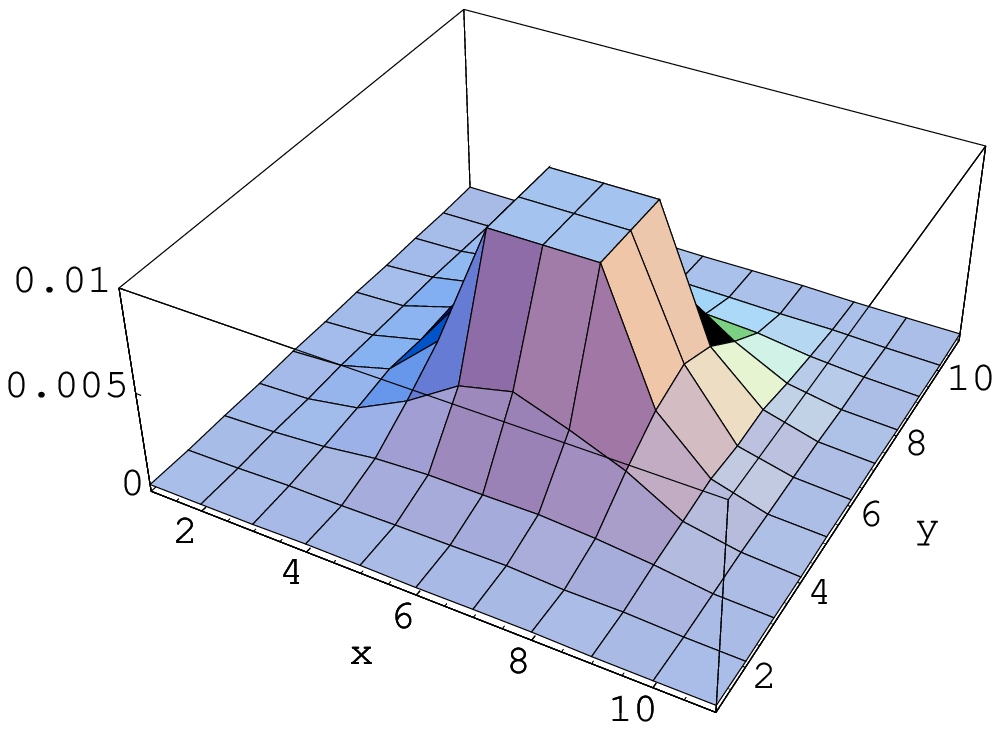}
\vglue 0.4cm
\epsfxsize=3.1in
\epsfysize=2.9in
\epsffile{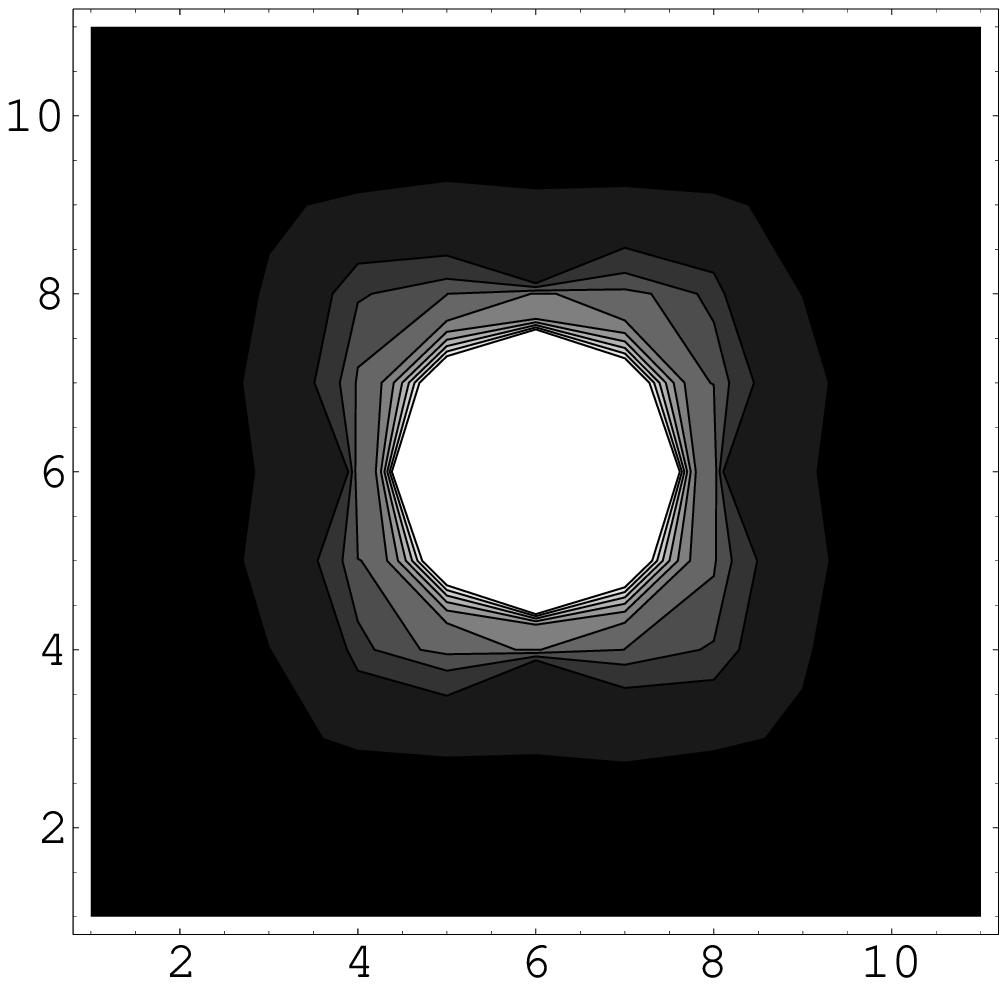}
\vglue 0.2cm
\caption{~Decay~of~$C({\bf{r}})$~for~Hamiltonian(1)~with $U/t =0,W/t = 7, N = 10$ and
$\nu = 1/4$ (50 fermions on a $10 \times 10$ lattice),
averaged over 16 $W$ values.
The upper part(a) shows the decay in  3D form for the interval
$0 \leq C({\bf{r}}) \leq 0.01$, the lower part(b) is
a contour plot of the same data.
}
\label{fig3ab}
\end{figure}

The dependence of $C({\bf{r}})$ on ${\bf{r}}$, shown in Figs. 3a,b for
the 2D disordered, non-interacting Hubbard model model ($U/t = 0$), 
also clearly shows localisation
of the ground state eigenfunction. The decay 
(as seen from the contour plot Fig. 3b)
is approximately symmetric
in ${\bf{r}}$ which is due to averaging over different disorder
realizations.
Hence, it should be useful to study this quantity 
also in the interacting case, to clarify the 
ground state properties.

\begin{figure}
\epsfxsize=3.1in
\epsfysize=2.9in
\epsffile{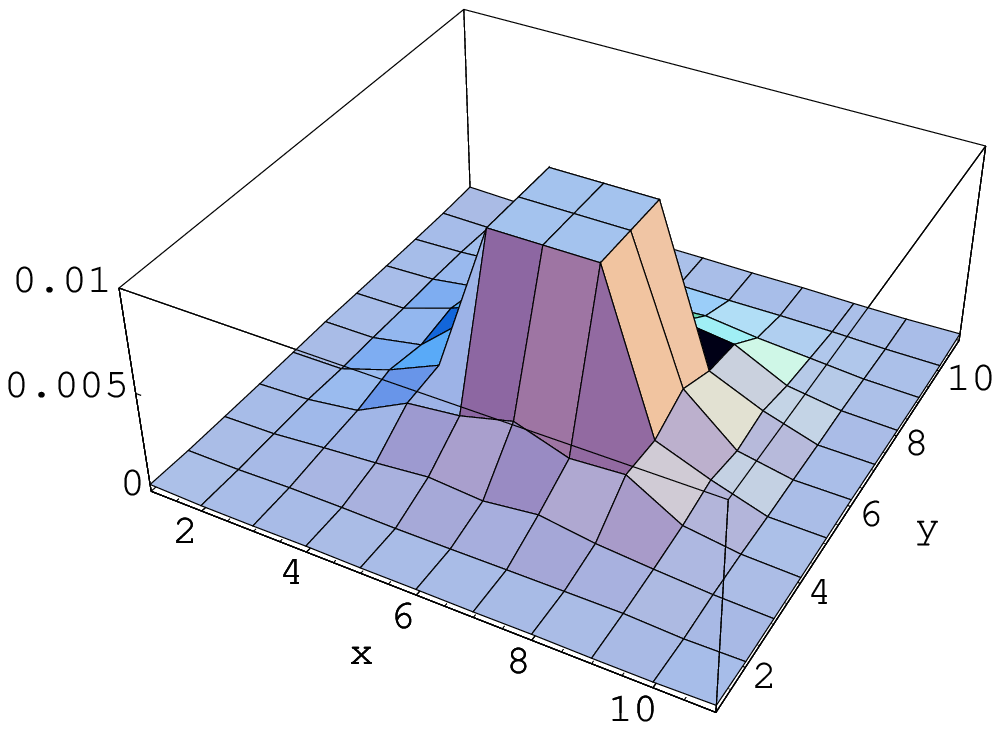}
\vglue 0.4cm
\epsfxsize=3.1in
\epsfysize=3.0in
\epsffile{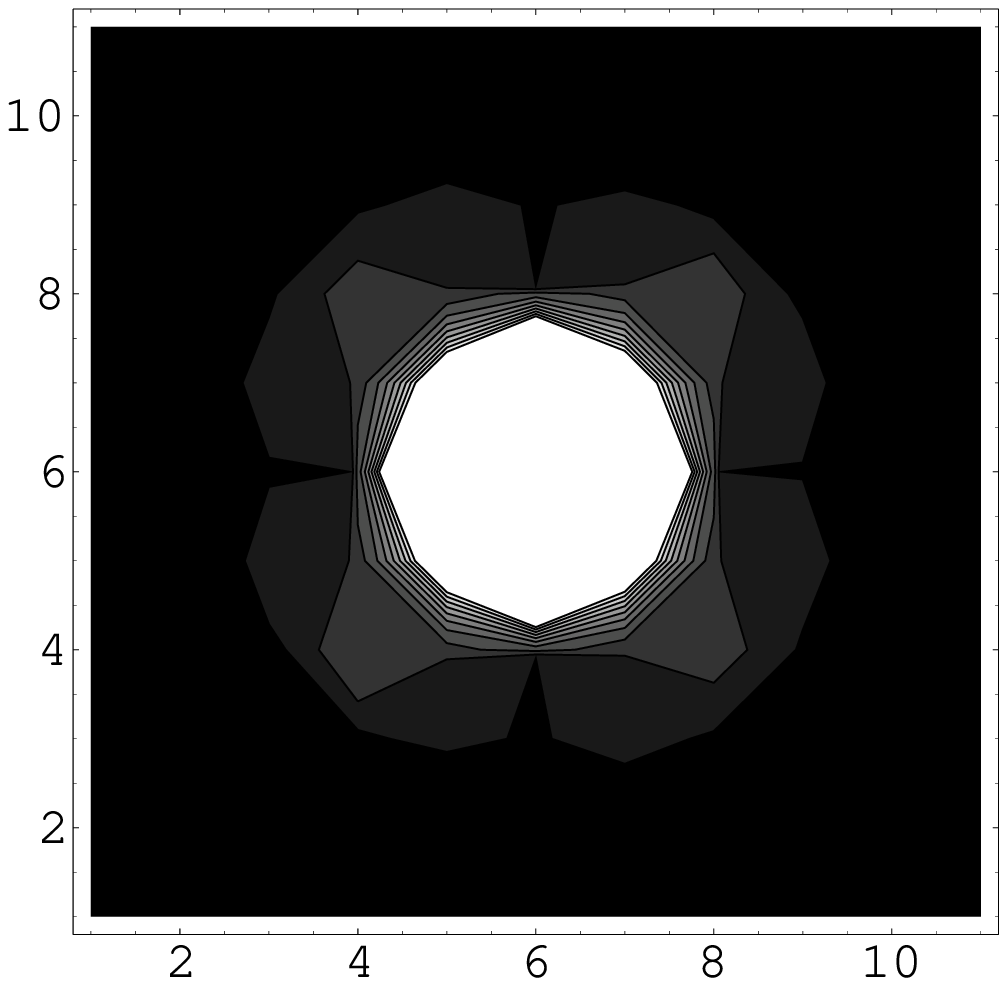}
\vglue 0.2cm
\caption{
Same as in Fig. 3a,b with $U/t = 6$ and the same disorder realisations.
}
\label{fig4ab}
\end{figure}

The behaviour of  $C({\bf{r}})$ as a function of  ${\bf{r}}$, 
for relatively strong interaction strength, ($U = 6t$)
is shown in Figs. 4a,b. The comparison with the non-interacting
case (Figs. 3a,b) clearly shows that even such a strong interaction
produces only a slight change in the correlation function.
We observe similar behaviour for
other disorder and interaction strengths ( $5 \leq W/t \leq 10$,
$ 0 < U/t \leq 6 $, $\nu$ = 1/4, 1/8, results not presented here).
This, in our opinion, provides direct evidence that even in 
the presence of strong interactions, the
ground state of the system remains localised.  
This conclusion is futher supported by the data for the
direction averaged correlation function, $\bar{C}(r)$ presented in
Fig. 5.  In fact, this direction average further smoothens 
fluctuations due to disorder.
Indeed, even the introduction of relatively strong interactions 
($U/t$ = 6) affects this
function very weakly.

\begin{figure}
\epsfxsize=3.1in
\epsfysize=3.0in
\epsffile{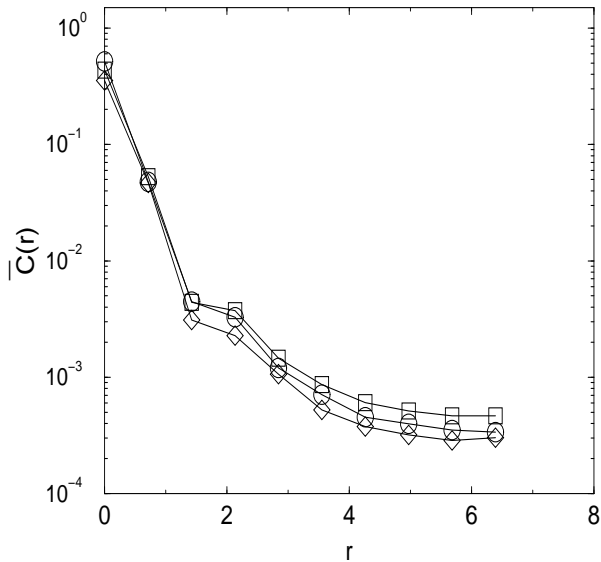}
\vglue 0.2cm
\caption{
 The decay of the direction averaged correlation function
 $\bar{C}(r)$ vs. $r$ 
 for the Hamiltonian (1) with $U/t$  = 0 (circles), 2 (squares),
 6 (diamonds) for  $W/t = 7, N = 10$ and 
 filling $\nu = 1/4$ (50 fermions on a $10 \times 10$ lattice),
 averaged over the same 16 disorder realisations.
}
\label{fig5}
\end{figure}

As an alternative test for localisation, we use another,  more
indirect method, similar to the approach presented in 
Ref.\cite{pichard2}. As discussed in \cite{pichard2}, we
vary the amplitude of on-site disorder  $\epsilon_i$ for
sites $i$ along one vertical line of the 
square lattice, as $\epsilon_i \rightarrow a \times \epsilon_i $
with $a$ = 1.1 and 1.3, corresponding to 10\% and 30\% 
change in disorder. We then study the charge density difference
$\delta\rho_x$
produced by this perturbation, as a function of distance $x$ from
the original line. We average over all sites with the same $x$
and additionally average  $\log |\delta\rho_x| $
over 16 disorder realisations.  The comparison of data for 
10\% and 30\% variation shows that we are in the linear 
response regime.  The  results  are presented  in Fig. 6.
For $U/t = 0$, the response function shows a sharp drop
from the initial peak followed by a slower decay at longer
distances. This behaviour is qualitatively similar to the
decay of the correlation function (Fig. 2, Fig. 5), for the
same physical reasons as analysed above. Introducing interactions
doesn't at all affect the main structure of the curve which
drops very quickly from the centre by more than one
order of magnitude. We interpret this as a sign of a localised
ground state.
At the same time, we note that there is a slight difference
introduced by interactions at the tails of the response
functions. However this corresponds to a density variation
less than 0.1\%, which is at the limit of the accuracy 
of our calculation. In the light of the ensemble of data, we
conclude that the ground state in the presence of interactions
remains localised.

\begin{figure}
\epsfxsize=3.1in
\epsfysize=3.0in
\epsffile{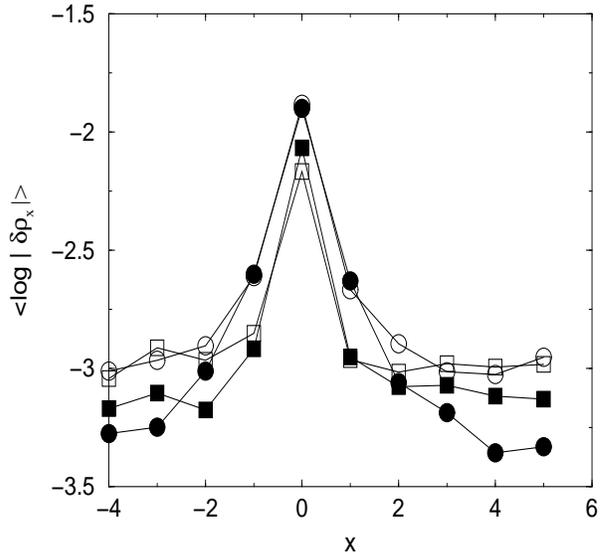}
\vglue 0.2cm
\caption{ Behaviour of $\langle \log |\delta\rho_x| \rangle$ with $x$
for 
$U/t$ = 0 (filled
symbols), 2 (open symbols) and 
$W/t$ = 2 (squares) and  7 (circles)  with $N$ = 10,
$\nu$ = 1/4 and $a$ = 1.1, averaged over the same 16 realisations.
}
\label{fig6}
\end{figure}

In a sense the Hubbard repulsion leads to local rearrangements of
charge and does not seem to influence the long-distance properties
of the system.  This is clearly illustrated in Figs. 7a,b
where the interaction  leads to a more homogeneous charge
distribution (note the change in the vertical scale) but 
doesn't drastically change the global profile.

This point of view is further borne out by the 
data for the inverse participation ratio (IPR), $\xi$,
presented in Fig. 8. The IPR, defined as 
$\xi =  (\sum_i n_i)^2 /\langle (N_p \sum_i ( n_i)^2 ) \rangle$, gives the
average number of sites visited per particle.  
$\xi$ is bounded from above at fixed filling, 
with $\xi \leq \xi_{max} =
(2\nu)^{-1}$, corresponding to the weak disorder limit
and from below with
 $\xi \geq \xi_{min} = 0.5$ in the strongly localised limit.
We note that $C(0)\times\xi = 2\nu$.
We have studied the IPR as a function of
system size and interaction strength at 1/4 (Fig. 8) 
and 1/8 (Fig. 9) filling. At $U/t = 0$, the IPR naturally increases
with decreasing disorder as states become more extended.
At strong disorder ($W/t \geq 5$)  $\xi$ is not sensitive to the system
size since the states are localised and at fixed filling the
IPR is counted per particle. The Hubbard interaction 
smoothly increases the IPR but does not introduce a
qualitative change. This corresponds to the local 
reorganisation of charge introduced by $U$
leading to a more homogeneous charge density 
distribution, as discussed above.
The data presented for 1/8 filling in Fig. 9 show 
qualitatively similar behaviour. However, the size
variation is more restricted in this case ( $ N \leq 10$ in 
our studies).

\begin{figure}
\epsfxsize=3.1in
\epsfysize=3.0in
\epsffile{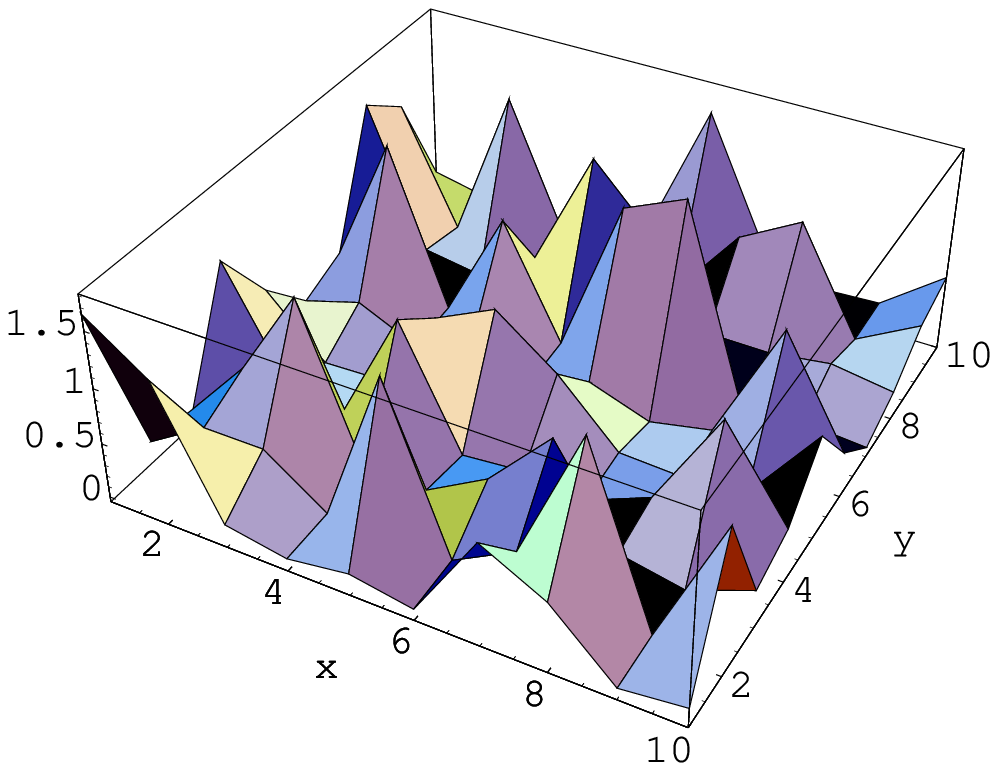}
\vglue 0.2cm
\epsfxsize=3.1in
\epsfysize=3.0in
\epsffile{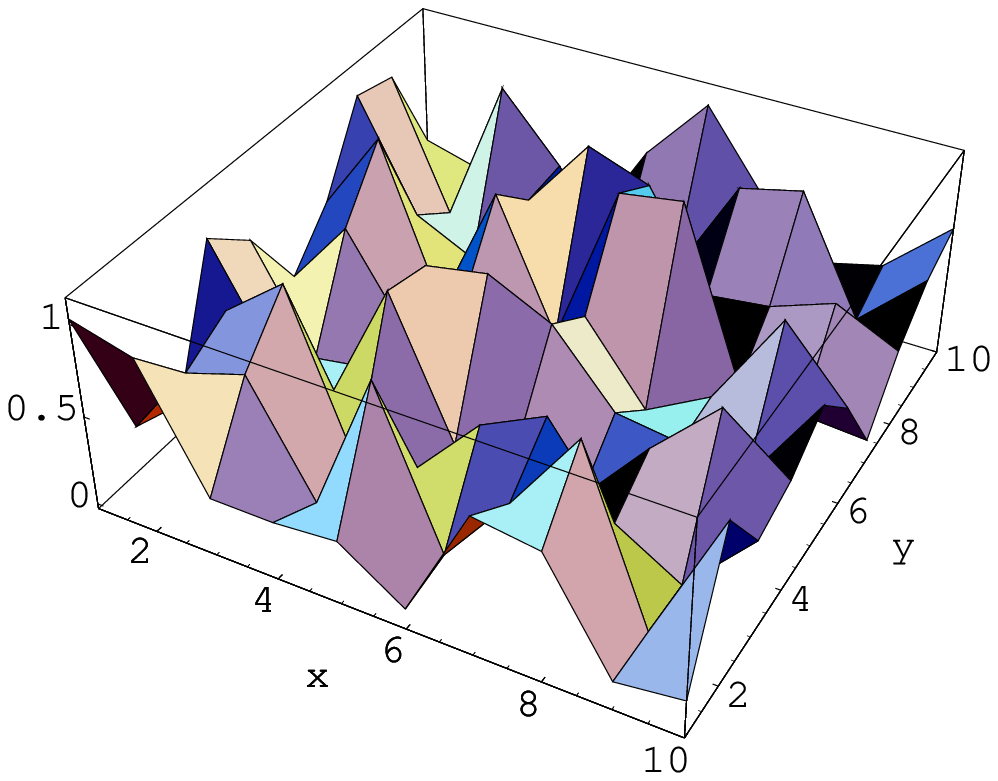}
\vglue 0.2cm
\caption{ Charge density ($n_i$) at site $i$ for 
$U/t$ = 0 (upper figure (a)) and 6 (lower figure (b)) 
for a $10 \times 10$ lattice, $W/t$ = 7, $\nu$=1/4.
}
\label{fig7}
\end{figure}

\begin{figure}
\epsfxsize=3.1in
\epsfysize=3.0in
\epsffile{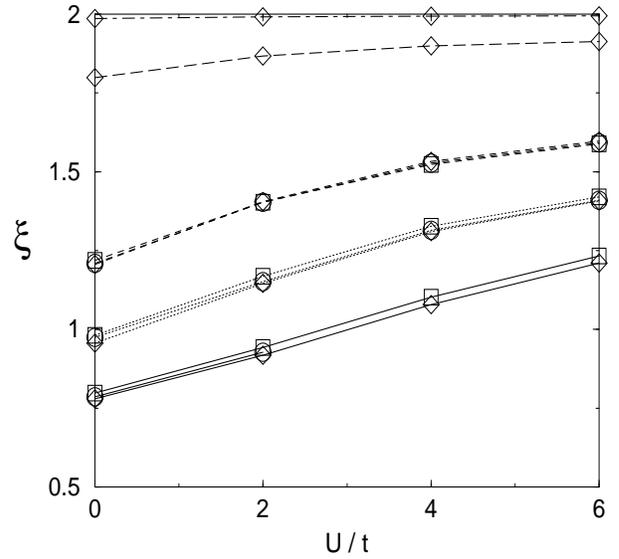}
\vglue 0.2cm
\caption{
IPR ($\xi$) vs. $U/t$ for system sizes $10 \times 10$ (circles), $8 \times 8$ (squares) and
$ 6 \times 6$ (diamonds) and increasing disorder strength from top
to bottom, $W/t$ = 0.5 (dot-dashed), 2 (long-dashed),
5 (dashed), 7 (dotted) and 10 (solid) lines, at quarter filling,
averaged over the same 16 disorder realisations.
}
\label{fig8}
\end{figure}

\begin{figure}
\epsfxsize=3.1in
\epsfysize=3.0in
\epsffile{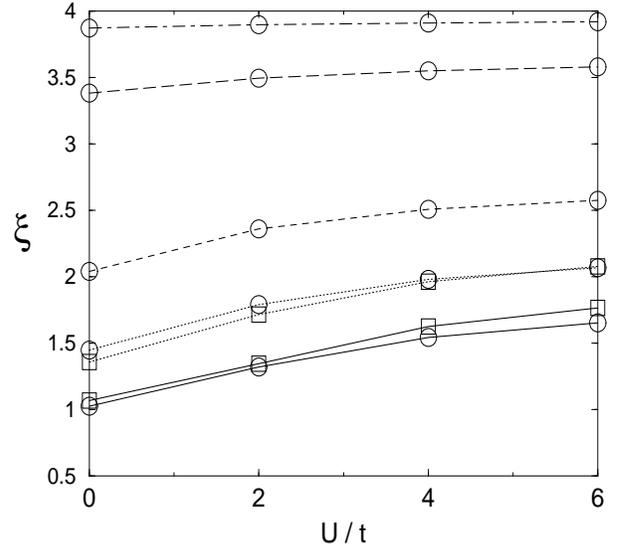}
\vglue 0.2cm
\caption{
Same as in Fig. 8 for 
for system sizes $8 \times 8$ (circles) and $4 \times 4$ (squares)
at 1/8 filling.
}
\label{fig9}
\end{figure}

At this point, it is interesting to compare our studies 
with a recent 
finite temperature QMC study 
of a similar model\cite{trivedi}. These authors considered
the Hubbard model on a square lattice, with off-diagonal
disorder, in contrast to our study. This is due to the 
fact that the method used, a finite temperature QMC method,
suffers from a severe sign problem in the presence of
diagonal disorder. For off-diagonal disorder, the 
situation becomes better, but the problem persists,
restricting the lowest accessible temperatures.
Their studies of several physical characteristics, 
including the conductivity (obtained
by approximate analytic continuation of the
imaginary time Green function), indicate the presence of an
interaction induced metal-insulator transition in their
model. However, this method is not adapted to analysis
of the ground state properties. Our results are not in
direct contradiction with this study, since it is fully 
possible that the ground state remains localised, while
the low-lying excited states become delocalised. Such
a situation has been observed in numerical studies 
of spinless fermions with Coulomb interactions on a 
2D lattice with disorder\cite{song}. Our result
directly demonstrates the localised nature of the
ground state even in the presence of strong interactions.
This is important in the framework of general studies 
of zero-temperature
quantum phase transitions.

\section{Conclusions}

We have used the projector quantum Monte Carlo (PQMC) method
to study the ground state of the 2D disordered Hubbard model.
This method allows us to study systems of up to 50 spin-1/2 fermions 
on a $10 \times 10$ lattice, for interaction strengths $U/t$
up to 6 and a broad interval of disorder  strengths.
The comparison of several properties  in the absence and
presence of the Hubbard interaction allows us to
conclude that interactions lead to local rearrangements 
of charge but do not destroy the localised
structure of the ground state, within the range of parameter
values studied.

These results indicate that short-range interactions are
probably insufficient to bring about a quantum phase transition
in the ground state of
this system. Thus, it becomes important to consider the
effect of long-range Coulomb interactions for electrons on a
disordered lattice. 

We thank M-B. Lepetit for useful discussions and the IDRIS, Orsay,
for allocation of resources on their supercomputers.


\vskip -0.5cm
\begin{thebibliography}{99}
\bibitem{and58} P.~W.~Anderson, Phys. Rev. {\bf 109}, 1492 (1958).
\bibitem{and79} E.~Abrahams, P.~W.~Anderson, D.~C.~Licciardello,
        and T.~V.~Ramakrishnan,  Phys. Rev. Lett. {\bf 42}, 673 (1979).
\bibitem{lee} P.~A.~Lee, and T.~V.~Ramakrishnan, Rev. Mod. Phys. {\bf 57},
           287 (1985).
\bibitem{krav94} S.~V.~Kravchenko, G.~V.~Kravchenko, J.~E.~Furneaux,
        V.~M.~Pudalov, and M.~D'Iorio, Phys. Rev. B  {\bf 50},
        8039 (1994); S.~V.~Kravchenko, W.~E.~Mason, G.~E.~Bowker, 
        J.~E.~Furneaux, V.~M.~Pudalov, and M.~D'Iorio, 
        Phys. Rev. B  {\bf 51}, 7038 (1995).
\bibitem{popovic} D.~Popovi\'c, A.~B.~Fowler, and S.~Washburn,
        Phys. Rev. Lett. {\bf 79}, 1543 (1997).
\bibitem{canada} P.~T.~Coleridge, R.~L.~Williams, Y.~Feng,
        and P.~Zawadzki, Phys. Rev. B {\bf 56}, R12764 (1997).
\bibitem{yael} Y.~Hanein, D.~Shahar, J.~Yoon, C.~C.~Li,
        D.~C.~Tsui, and H.~Shtrikman, Phys. Rev. B
        {\bf 58}, R7520 (1998).
\bibitem{alex} M.~Y.~Simmons, A.~R.~Hamilton, M.~Pepper,
        E.~H.~Linfield, P.~D.~Rose, and D.~A.~Ritchie,
        Phys. Rev. Lett. {\bf 80}, 1292 (1998).
\bibitem{simonian} D.~Simonian, S.~V.~Kravchenko, M.~P.~Sarachik,
        and V.~M.~Pudalov, Phys. Rev. Lett. {\bf 79}, 2304 (1997).
\bibitem{alar} B.~L.~Altshuler and A.~G.~Aronov, in:
        Electron-Electron Interactions in Disordered Systems, eds.
        A.~L.~Efros and M.~Pollak (North-Holland, Amsterdam, 1985), p.1.
\bibitem{efros} A.~L.~Efros and B.~I.~Shklovskii, J. Phys. C {\bf 8}, L49 
        (1975);  see also 
        {\it Electron--Electron Interactions in Disordered Systems}, 
        eds.  A.~L.~Efros and M. Pollak, (North--Holland, Amsterdam, 1985),
        p. 409.
\bibitem{ds94} D.~L.~Shepelyansky, Phys. Rev. Lett. 
        {\bf 73}, 2607 (1994).
\bibitem{mirlin} A.~D.~Mirlin, Phys. Rep. {\bf 326}, 259 (2000).
\bibitem{benen} G.~Benenti, X.~Waintal and J.-L.~Pichard,
       	Phys. Rev. Lett. {\bf 83}, 1826 (1999).
\bibitem{poilblanc} G.~Bouzerar and D.~Poilblanc, J. Phys. I France 
        {\bf 7}, 877 (1997). 
\bibitem{vojta} T.~Vojta, F.~Epperlein, and M.~Schreiber, 
	Phys. Rev. Lett. {\bf 81}, 4212 (1998).
\bibitem{hfgap} G.~Benenti, X.~Waintal, J-L.~Pichard and D.~L.~Shepelyansky,
	pre-print no. cond-mat/0003208.
\bibitem{song} P.~H.~Song and D.~L.~Shepelyansky, pre-print no. cond-mat/9904229;
        Ann. Phys. (Leipzig) {\bf 8}, 665 (1999).
\bibitem{trivedi} P.~J.~H.~Denteneer, R.~T.~Scalettar and N.~Trivedi,
	Phys. Rev. Lett. {\bf 83}, 4610 (1999).
\bibitem{krav3}S. V. Kravchenko, T. M. Klapwijk,
        Phys. Rev. Lett. {\bf 84}, 2909 (2000).
\bibitem{pqmc} M. Imada and Y. Hatsugai, J. Phys. Soc. Jpn.
        {\bf 58},  3752 (1989).
\bibitem{bs1} B.~Srinivasan, S.~Ramasesha and H.~R.~Krishnamurthy,
	Phys. Rev. B {\bf 54} 1692 (1996).
\bibitem{pichard2} X.~Waintal, G.~Benenti, and J.-L.~Pichard,
        Europhys. Lett. 49, 466 (2000).





\end{thebibliography}
\end{document}